\documentclass[]{JHEP3}

\usepackage{graphicx}
\usepackage{amsmath}
\def\text{\textrm}
\title{Notes on Ghost Dark Energy}

\author{Rong-Gen Cai, Zhong-Liang Tuo, Hong-Bo Zhang\\
    Key Laboratory of Frontiers in Theoretical Physics,
    Institute of Theoretical Physics, Chinese Academy of Sciences,
    P.O. Box 2735, Beijing 100190, China\\
    E-mail:
    \email{cairg@itp.ac.cn}
    \email{tuozhl@itp.ac.cn}
    \email{hbzhang@itp.ac.cn}}
\author{Qiping Su\\
    Department of Physics, Hangzhou Normal University, Hangzhou, 310036,
    China\\
    E-mail:
    \email{sqp@hznu.edu.cn}}

\received{\today}       %%
\accepted{\today}       %% These are for published papers.

\abstract{We study a phenomenological dark energy model which is
rooted in the Veneziano ghost of QCD. In this dark energy model, the
energy density of dark energy is proportional to Hubble parameter
and the proportional coefficient is of the order $\Lambda^3_{QCD}$,
where $\Lambda_{QCD}$ is the mass scale of QCD.  The universe has a
de Sitter phase at late time and begins to accelerate at redshift
around $z_{acc}\sim0.6$. We also fit this model and give the
constraints on model parameters,  with current observational data
including SnIa, BAO, CMB, BBN and Hubble parameter data. We find
that the squared sound speed of the dark energy is negative, which
may cause an instability. We also study the cosmological evolution
of the dark energy with interaction with cold dark matter.}

\begin{document}

%\maketitle  IS IGNORED %%%%%%%%%%%

\section{Introduction}

It has been more than ten years since our universe was found to be
in accelerating expansion \cite{Riess:1998cb}. A new energy
component of the universe, called dark energy (DE), is needed to
explain this acceleration. The simplest model of DE is the
cosmological constant, which is a key ingredient in the
$\Lambda\text{CDM}$ model. Although the $\Lambda\text{CDM}$ model is
consistent very well with all observational data, it faces with the
fine tuning problem \cite{finetunning}. Plenty of other DE models
have also been proposed
\cite{Copeland:2006wr,Caldwell:1997mh,Steinhardt:1999nw,Capozziello:2003tk,Li:2004rb,Nojiri:2005sr,Nojiri:2003ft},
but almost all of them explain the acceleration expansion either by
introducing new degree(s) of freedom or by modifying gravity.

Recently a new DE model, so-called Veneziano ghost DE, has been
proposed \cite{Urban,Ohta:2010in}. The key ingredient of this new
model is that the Veneziano ghost, which is unphysical in the usual
Minkowski spacetime QFT, exhibits important physical effects in
dynamical spacetime or spacetime with non-trivial topology.
Veneziano ghost is supposed to exist for solving the $U(1)$ problem
in low-energy effective theory of QCD~
\cite{Witten,Veneziano,RST,NA,KO}. The ghost has no contribution to
the vacuum energy density in Minkowski spacetime, but in curved
spacetime it gives rise to a small vacuum energy density
proportional to $\Lambda_{QCD}^{3}H$, where $\Lambda_{QCD}$ is QCD
mass scale and $H$ is Hubble parameter. This small vacuum energy
density expects to play some role in the evolution of the universe.
 Because this model is totally embedded in
standard model and general relativity, one needs not to introduce
any new parameter, new degree of freedom or to modify gravity. With
$\Lambda_{QCD}\sim100MeV$ and $H\sim10^{-33}eV$,
$\Lambda_{QCD}^{3}H$ gives the right order of observed DE energy
density. This numerical coincidence is impressive and also means
that this model gets rid of fine tuning problem
\cite{Urban,Ohta:2010in}. Actually, several authors have already
suggested DE model with energy density proportional to
$\Lambda_{QCD}^{3}H$ in different physical contexts
\cite{Bjorken:2001pe,Schutzhold:2002pr,Bjorken:2004an,Klinkhamer:2007pe,
Klinkhamer:2008ns,Klinkhamer:2009nn,Zeldovich:1967gd,Zimdahl:2010ae,Nojiri:2005sr}
such as QCD trace anomaly, gluon condensate of quantum
chromodynamics, modifying gravity and so on.

In this work, we investigate the phenomenological model with energy
density of DE $\rho_{DE}$ proportional to Hubble parameter $H$. We
study the cosmological evolution of the DE model with/without
interaction between the DE and dark matter. We analytically and
numerically compute some quantities such as  scale factor $a$,
$\rho_{DE}$, squared adiabatic speed of sound $c_s^2$ and so on.
Also we fit this model with current observational data and give
constraints on the model parameters.

The note is organized as follows. In section 2 we study the
dynamical evolution of the DE model. In section 3, we fit this model
with current observational data and discuss the fitting results. The
data used are Union II SnIa sample \cite{Amanullah:2010vv}, BAO data
from SDSS DR7 \cite{Percival:2009xn}, CMB data $(R,\, l_{a},\,
z_{*})$ from WMAP7 \cite{Komatsu:2010fb}, 12 Hubble evolution data
\cite{Simon:2004tf,Gaztanaga:2008xz} and Big Bang Nucleosynthesis
(BBN) \cite{Serra:2009yp,Burles:2000zk}. In section 4, we calculate
$c_s^2$ and find it is always negative.  Negative $c_s^2$ may give
rise to the instability problem. On the one hand, to try to avoid
this problem, on the other hand, to further study the ghost DE
model,  we introduce the interaction between DE and cold dark matter
(CDM), and study the dynamical evolution in this case. We summary
our work and give some discussions in section 5.

%%%%%%%%%%%%%%%%%%%%%%%%%%%%%%%%%%%%%%%%%%%%%%%%%%%%%%%%%%%%%%%%%%%%%%%%%%%%%%%%%%

\section{Dynamics of ghost dark energy}

To study the dynamics of the DE model, we  consider a flat FRW
universe with only two energy components, CDM and DE  and neglect
radiation and baryon temporarily in this section.  We will include
the radiation and baryon when fit the model with observational data
in section 4.

In this ghost DE model,  the energy density of DE is given by
$\rho_{DE}=\alpha H$, where $\alpha$ is a constant with dimension
$[energy]^{3}$, and roughly of order of $\Lambda_{QCD}^{3}$ where
$\Lambda_{QCD}\sim100\text{MeV}$ is QCD mass scale.  Arming with
this DE density, the Friedman equation reads\begin{equation}
H^{2}=\frac{8\pi G}{3}\left(\alpha
H+\rho_{m}\right),\label{eq:friedman}\end{equation}
 where $\rho_{m}$ is energy density of CDM, whose continuity
 equation gives \begin{equation}
\dot{\rho}_{m}+3H\rho_{m}=0\;\Longrightarrow\;\rho_{m}=\rho_{m0}a^{-3}.
\label{eq:cdm_conservation}\end{equation}
 We have set $a_{0}=1$ and the subscript $0$ stands for the present value of some quantity.
From (\ref{eq:friedman}) and (\ref{eq:cdm_conservation}), we can
obtain the Raychaudhuri equation
\begin{equation}
\dot{H}+H^{2}=-\frac{4\pi
G}{3}\left[-\rho_{DE}\left(\frac{\dot{\rho}_{DE}}
{H\rho_{DE}}+2\right)+\rho_{m}\right].
  \label{eq:Raychaudhuri}
\end{equation}
 Solve the Friedman equation, we have
  \begin{equation}
H_{\pm}=\frac{4\pi G}{3}\alpha\pm\sqrt{\left(\frac{4\pi
G}{3}\alpha\right)^{2} +\frac{8\pi G}{3}\rho_{m0}a^{-3}}.
\label{eq:hubble}\end{equation} There are two branches, $H_{+}$
represents an expansion solution, while $H_{-}$  a contraction one.
We neglect the latter since it goes against the observation, and for
simplicity, write $H_{+}$ as $H$ in what follows.

To facilitate our discussion, we define a characteristic scale
factor $a_{*}$ \[ a_{*}\equiv\left(\frac{12\rho_{m0}}{8\pi
G\alpha^{2}}\right)^{\frac{1}{3}}=\left(4\frac{\Omega_{m0}}{\Omega_{DE0}^{2}}\right)^{\frac{1}{3}}
=10^{2}\left(36\Omega_{m0}\frac{\text{MeV}^{6}}{\alpha^{2}}\right)^{\frac{1}{3}},\]
where we have taken $H_{0}=10^{-33}eV$, $\Omega_{m0}$ and
$\Omega_{DE0}$ are the dimensionless energy density of CDM and DE,
respectively. One can see shortly that actually $a_{*}$ is the
transition point when the universe transits from the dust phase to a
de Sitter phase. If we assume $\Omega_{m0}=\frac{1}{4}$ and
$\Omega_{DE0}=\frac{3}{4}$, by definition, we get roughly
$a_{*}\sim1$, which means that the transition occurs just at
present. Therefore, we will take $a_{*}=1$ throughout this section.
And especially $\alpha\sim(10\text{MeV})^3$ if $a_{*}\sim1$.

At early epoch $a\ll a_{*}\sim1$ , the $a^{-3}$ term dominates in
(\ref{eq:hubble}), the Hubble parameter behaves like $H\sim
a^{-\frac{3}{2}}$,  which means that the universe is in a dust
phase. While at late epoch $a\gg a_{*}\sim1$, the $a^{0}$ term
dominates in (\ref{eq:hubble}), as a result, $H=\text{const}$, which
says that the universe enters a de Sitter phase at late time. Here
$a_{*}$ is the transition point between these two phases as
mentioned above.

\begin{figure}
\begin{centering}
\includegraphics{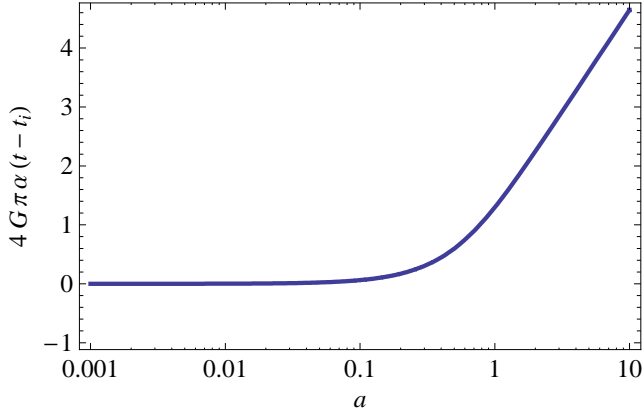}
\par\end{centering}
\caption{$4\pi G\alpha(t-t_{i})\sim a$, where  $a_{*}=1$
\label{fig:tofa}}

\end{figure}

We can solve (\ref{eq:hubble}) analytically as
\begin{eqnarray*}
4\pi G\alpha\left(t-t_{i}\right)  & = &
-x^{3}+x^{3}\sqrt{1+x^{-3}}+\frac{3}{2}\ln
x+\ln\left(1+\sqrt{1+x^{-3}}\right),\end{eqnarray*}
 where $x=\frac{a}{a_{*}}$ and $t_{i}$ is the initial time when $a(t_{i})=0$.
At early time $x\ll1$, $4\pi
G\alpha\left(t-t_{i}\right)=2x^{\frac{3}{2}}$; while at late time
$x\gg1$, $4\pi G\alpha\left(t-t_{i}\right)=\frac{3}{2}\ln x$. These
asymptotic behaviors agree with the previous argument. The numerical
relation $t\sim a$  is plotted in Figure~\ref{fig:tofa}.

\begin{figure}
\begin{centering}
\includegraphics{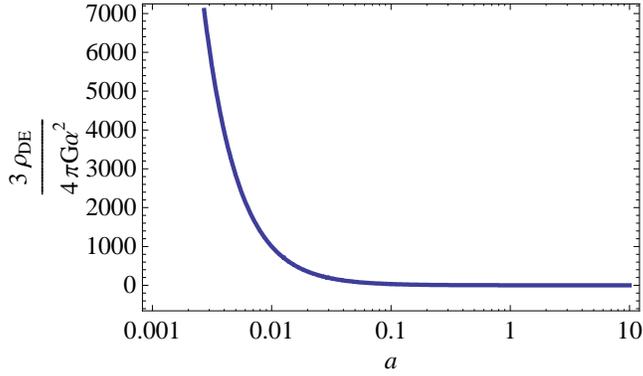}
\par\end{centering}
\caption{$\frac{3H}{4\pi G\alpha}\sim a$, where
$a_{*}=1$\label{fig:hofa}}
\end{figure}

\begin{figure}
\begin{centering}
\includegraphics{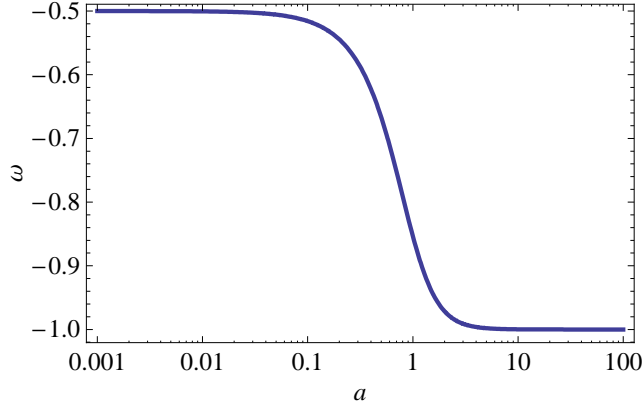}
\par\end{centering}
\caption{$\omega\sim a$, where  $a_{*}=1$\label{fig:wofa}}
\end{figure}

Using (\ref{eq:hubble}) and the definition of $\rho_{DE}$, the
energy density of DE is\[ \rho_{DE}=\alpha H=\frac{4\pi
G}{3}\alpha^{2}\left[1+\sqrt{1+\left(\frac{a_{*}}{a}\right)^{3}}\right].\]
The behavior of  $\rho_{DE}$ in terms of  $a$  is shown in Figure
~\ref{fig:hofa}.
The equation of state (EoS) of the ghost DE is given by
\begin{eqnarray}
\omega & \equiv & -\frac{1}{3}\frac{\dot{\rho}_{DE}}{H\rho_{DE}}-1=\frac{1}
{2}\left(\frac{a_{*}}{a}\right)^{3}\left(\frac{1}{\sqrt{1+\left(\frac{a_{*}}
{a}\right)^{3}}}-\frac{1}{1+\sqrt{1+\left(\frac{a_{*}}{a}\right)^{3}}}\right)-1\label{eq:w_a} \\
 & = & \left\{
 \begin{array}{ll}
-\frac{1}{2} & \qquad a\ll a_{*}\\
-1 & \qquad a\gg a_{*}
\end{array}\right..\nonumber
\end{eqnarray}
 From the asymptotic behavior of $\omega$, we can see that the DE acts
like a cosmological constant at late time. We plot the relation
$\omega\sim a$  in Figure~\ref{fig:wofa}. From the figure, we can
see that $\omega$ can never cross $-1$, which is similar to the
behavior of quintessence. We will show that this behavior can be
altered in the presence of interaction between DE and CDM in section
4. $\omega$ varies from $-\frac{1}{2}$ at early time to $-1$ at late
time,  which is similar to freezing quintessence model
\cite{Caldwell:2005tm}. We can also find that $\omega$ has a sharp
variation round $a=a_{*}$. It is easy to understand  if we rewrite
the expression of $\omega$ as
\begin{equation}
3(1+\omega)=-\frac{\dot{\rho}_{DE}}{H\rho_{DE}}=-\frac{\dot{H}}{H^{2}}
=\left(H^{-1}\right)^{\cdot},\label{eq:w_h-1}\end{equation}
 from this equation, we can see that in this model the EOS of DE tightly
relates to the variation of Hubble parameter, which is quite
different in different phases of the universe. For instance, in the
dust phase, $\left(H^{-1}\right)^{\cdot}\sim\frac{3}{2}$; while in
the  de Sitter phase, $\left(H^{-1}\right)^{\cdot}\sim0$. Therefore,
there will be a jump from $-\frac{1}{2}$ to $-1$ when the universe
transits from the dust phase to the de Sitter phase.

One can easily show that the EoS of the ghost DE model is
$\omega=-\frac{1}{\Omega_{m}+1}$. The value  $\omega_{0}$ at present
is
\begin{eqnarray} \omega_{0}\left(a=1\right)
& = & \frac{1}{\Omega_{DE0}-2}=-\frac{1}{\Omega_{m0}+1},
\label{eq:w_Omegam}
\end{eqnarray}
 where we have used $a_{*}=\left(4\frac{\Omega_{m0}}{\Omega_{DE0}^{2}}\right)^{\frac{1}{3}}$
in the first equality. This relation is important and helpful to
understand the fitting results in section 3.

Also from Raychaudhuri equation (\ref{eq:Raychaudhuri}), we can get
the total equation of state of the universe
\begin{equation}
\omega_{tot}=-1-\frac{2}{3}\frac{\dot{H}}{H^{2}}=1+2\omega= \left\{
\begin{array}{ll}
0 & \qquad a\ll a_{*}\\
-1 & \qquad a\gg
a_{*}.\end{array}\right.\label{eq:wtot_a}\end{equation}
$\omega_{tot}$ decreases monotonically from $0$ to $-1$, which means
that the expansion of the universe switches from deceleration at
early epoch to acceleration at late epoch. The conversion occurs at
$a_{acc}$ when $\omega_{tot}\left(a_{acc}\right)=-\frac{1}{3}$. From
(\ref{eq:w_a}) and (\ref{eq:wtot_a}), we can calculate
$a_{acc}=\frac{a_{*}}{2}\sim0.5$ which means that the universe begin
to accelerate at redshift $z_{acc}\sim1$.

%%%%%%%%%%%%%%%%%%%%%%%%%%%%%%%%%%%%%%%%%%%%%%%%%%%%%%%%%%%%%%%%%%%%%%5

\section{Data Fitting}

\subsection{Model}

In order to fit the model with current observational data, we
consider a more realistic model which includes DE, CDM, radiation
and baryon in a flat FRW universe in this section.  In this case,
the Friedman equation reads
\begin{eqnarray*}
H^{2} & = & \frac{8\pi G}{3}\left(\alpha
H+\rho_{DM}+\rho_{b}+\rho_{r}\right),
\end{eqnarray*}
which can be rewritten as
\begin{eqnarray*} E & \equiv &
\frac{H}{H_{0}}=\frac{1}{2}\Omega_{DE0}+\sqrt{\frac{1}{4}\Omega_{DE0}^{2}+\left(\Omega_{DM0}
+\Omega_{b0}\right)\left(1+z\right)^{3}+\Omega_{r0}\left(1+z\right)^{4}},\end{eqnarray*}
where $\Omega_{DM0},\Omega_{b0},\Omega_{r0}$ are present values of
dimensionless energy density for CDM, baryon and radiation,
respectively. $\Omega_{DE0}=\frac{8\pi G\alpha}{3H_{0}}$ is
dimensionless energy density of DE at present. Energy density of
baryon and CDM are always written together as
$\Omega_{DM0}+\Omega_{b0}=\Omega_{m0}$. Notice that
$\Omega_{DE0}+\Omega_{m0}+\Omega_{r0}=1$ since we assume a flat
universe. The energy density of radiation is the sum of those of
photons and relativistic neutrinos \[
\Omega_{r0}=\Omega_{\gamma0}\left(1+0.2271N_{n}\right),\]
 where $N_{n}=3.04$ is the effective number of neutrino species and $\Omega_{\gamma0}=2.469\times10^{-5}h^{-2}$
for $T_{cmb}=2.725K$ ($h={H_{0}}/{100}\, Mpc\cdot km\cdot s^{-1}$).

It is worth noticing  that from the definition of dimensionless
energy density of DE and flatness of our universe we can get an
important relation (see also (\ref{eq:H_OmegaDE})) \begin{equation}
\left(1-\Omega_{m0}\right)H_{0}=\frac{8\pi
G\alpha}{3}=\text{const},\end{equation}
 where we have neglected $\Omega_{r0}$, which is very small compared to $\Omega_{m0}$.
It means that parameters $\Omega_{m0}, h$ and $\alpha$ are closely
related, $\alpha$ can be expressed in terms of $\Omega_{m0}$ and
$h$. We will choose $\Omega_{m0},\, h$ and $\Omega_{b0}$ as free
parameters of the model in the following analysis. This relation
also infers that there exists a strong degeneracy between
$\Omega_{m0}$ and $h$, as shown in subsection 3.3.

\subsection{Sets of Observational data }

We fit our model by employing some observational data including
SnIa, BAO, CMB, Hubble evolution data and BBN.

The data for SnIa are the 557 Uion II sample \cite{Amanullah:2010vv}.
$\chi_{sn}^{2}$ for SnIa is obtained by comparing theoretical distance
modulus $\mu_{th}(z)=5\log_{10}[(1+z)\int_{0}^{z}dx/E(x)]+\mu_{0}$
($\mu_{0}=42.384-5\log_{10}h$) with observed $\mu_{ob}$ of supernovae:

\[
\chi_{sn}^{2}=\sum_{i}^{557}\frac{[\mu_{th}(z_{i})-\mu_{ob}(z_{i})]^{2}}{\sigma^{2}(z_{i})}.\]
 To reduce the effect of $\mu_{0}$, we expand $\chi_{sn}^{2}$ with
respect to $\mu_{0}$ \cite{Nesseris:2005ur} : \begin{equation}
\chi_{sn}^{2}=A+2B\mu_{0}+C\mu_{0}^{2}\label{eq:expand}\end{equation}
 where \begin{eqnarray*}
A & = & \sum_{i}\frac{[\mu_{th}(z_{i};\mu_{0}=0)-\mu_{ob}(z_{i})]^{2}}{\sigma^{2}(z_{i})},\\
B & = & \sum_{i}\frac{\mu_{th}(z_{i};\mu_{0}=0)-\mu_{ob}(z_{i})}{\sigma^{2}(z_{i})},\\
C & = & \sum_{i}\frac{1}{\sigma^{2}(z_{i})}.\end{eqnarray*}
 (\ref{eq:expand}) has a minimum as \[
\widetilde{\chi}_{sn}^{2}=\chi_{sn,min}^{2}=A-B^{2}/C,\]
 which is independent of $\mu_{0}$. In fact, it is equivalent to
performing an uniform marginalization over $\mu_{0}$, the difference
between $\widetilde{\chi}_{sn}^{2}$ and the marginalized $\chi_{sn}^{2}$
is just a constant \cite{Nesseris:2005ur}. We will adopt $\widetilde{\chi}_{sn}^{2}$
as the goodness of fit between theoretical model and SnIa data.

The second set of data is the Baryon Acoustic Oscillations (BAO)
data from SDSS DR7 \cite{Percival:2009xn}, the datapoints we use are
\[
d_{0.2}=\frac{r_{s}(z_{d})}{D_{V}(0.2)}\]
 and
\[
d_{0.35}=\frac{r_{s}(z_{d})}{D_{V}(0.35)},\]
 where $r_{s}(z_{d})$ is the comoving sound horizon at the baryon
drag epoch \cite{Eisenstein:1997ik}, and

\[
D_{V}(z)=\left[\left(\int_{0}^{z}\frac{dx}{H(x)}\right)^{2}\frac{z}{H(z)}\right]^{1/3}\]
 encodes the visual distortion of a spherical object due to the non
Euclidianity of a FRW spacetime. The inverse covariance matrix of
BAO is
\begin{eqnarray*} C_{M,bao}^{-1} & = & \left(\begin{array}{ccc}
30124 & -17227 \\
-17227 & 86977\end{array}\right).\end{eqnarray*}

The $\chi^2$ of the BAO data is constructed as:\[
\chi_{bao}^{2}=Y^{T}C_{M,bao}^{-1}Y,\] where
\[
Y=\left(\begin{array}{c}
d_{0.2}-0.1905\\
d_{0.35}-0.1097\end{array}\right).\]

The CMB datapoints we will use are ($R,l_{a},z_{*}$) from WMAP7 \cite{Komatsu:2010fb}.
$z_{*}$ is the redshift of recombination \cite{Hu:1995en}, $R$
is the scaled distance to recombination

\[
R=\sqrt{\Omega_{m}^{(0)}}\int_{0}^{z_{*}}\frac{dz}{E(z)},\]
 and $l_{a}$ is the angular scale of the sound horizon at recombination

\[
l_{a}=\pi\frac{r(a_{*})}{r_{s}(a_{*})},\]
 where $r(z)=\int_{0}^{z}dx/H(x)$ is the comoving distance and $r_{s}(a_{*})$
is the comoving sound horizon at recombination \[
r_{s}(a_{*})=\int_{0}^{a_{*}}\frac{c_{s}(a)}{a^{2}H(a)}da,\]
 where the sound speed $c_{s}(a)=1/\sqrt{3(1+\overline{R}_{b}a)}$
and $\overline{R}_{b}=3\Omega_{b}^{(0)}/4\Omega_{\gamma}^{(0)}$ is
the photon-baryon energy density ratio.

The $\chi^{2}$ of the CMB data is constructed as: \[
\chi_{cmb}^{2}=X^{T}C_{M,cmb}^{-1}X,\]
 where
\[
X=\left(\begin{array}{c}
l_{a}-302.09\\
R-1.725\\
z_{*}-1091.3\end{array}\right)\] and the inverse covariance matrix
\begin{eqnarray*} C_{M,cmb}^{-1} & = & \left(\begin{array}{ccc}
2.305 & 29.698 & -1.333\\
29.698 & 6825.270 & -113.180\\
-1.333 & -113.180 & 3.414\end{array}\right).\end{eqnarray*}

The fourth set of observational data is $12$ Hubble evolution data
from \cite{Simon:2004tf} and \cite{Gaztanaga:2008xz}. Its
$\chi_{H}^{2}$ is defined as

\[
\chi_{H}^{2}=\sum_{i=1}^{12}\frac{[H(z_{i})-H_{ob}(z_{i})]^{2}}{\sigma_{i}^{2}}.\]
 Note that the redshift of these data falls in the region $z\in(0,1.75)$.

The last set we will use is the Big Bang Nucleosynthesis (BBN) data
from \cite{Serra:2009yp,Burles:2000zk}, whose $\chi^{2}$ is\[
\chi_{bbn}^{2}=\frac{\left(\Omega_{b0}h^{2}-0.022\right)^{2}}{0.002^{2}}.\]

In summary, we have \[
\chi_{tot}^{2}=\widetilde{\chi}_{sn}^{2}+\chi_{cmb}^{2}+\chi_{bao}^{2}+\chi_{H}^{2}+\chi_{bbn}^{2}\]
and we assume uniform priors on all the parameters.

\subsection{Fitting Results}

\begin{figure}
\begin{centering}
\includegraphics[width=1\textwidth]{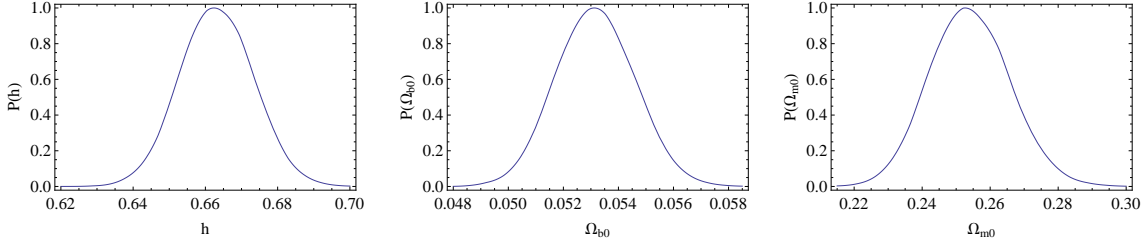}
\par\end{centering}

\caption{\label{fig:likelihood}1D marginalized distribution
probability of $h,\,\Omega_{b0}$ and $\Omega_{m0}$.}

\end{figure}

\begin{figure}
\begin{centering}
\includegraphics[width=0.7\textwidth]{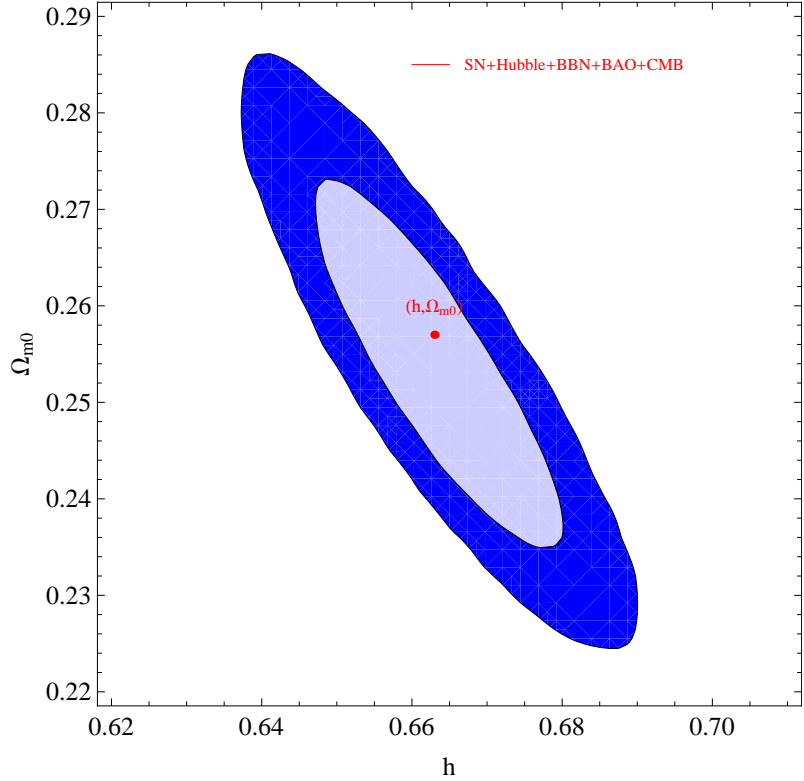}
\par\end{centering}

\caption{\label{fig:contour}$68\%$ and $95\%$ contour plot in
$\Omega_{m0}-h$ plane. The red dot in the figure stands for the
best-fit value.}

\end{figure}

The best-fit values and errors of parameters are summarized in
Table~\ref{tab:fit_result}. We also list the best-fit values of the
corresponding parameters in Table~\ref{tab:lcdm} for comparison. The
best-fit values of $\Omega_{m0}$ and $h$ are slightly smaller than
corresponding ones in the $\Lambda CDM$ model. In Figure
\ref{fig:likelihood}, we plot the 1D marginalized distribution
probability of each parameter. The 2D contour is plotted in Figure
\ref{fig:contour}, from which we can see that there exists a strong
correlation between $\Omega_{m0}$ and $h$ as we expected in
subsection 3.1.

\begin{table}
\begin{centering}
\begin{tabular}{|c|c|c|c|}
\hline parameter  & $\Omega_{m0}$  & $h$  &
$\Omega_{b0}$\tabularnewline \hline \hline
$\text{best-fit}_{+1\sigma,\,+2\sigma}^{-1\sigma,\,-2\sigma}$  &
$0.257_{+0.009,\,+0.020}^{-0.016,\,-0.026}$  &
$0.662_{+0.011,\,+0.021}^{-0.011,\,-0.019}$  &
$0.054_{+0.001,\,+0.002}^{-0.002,\,-0.004}$\tabularnewline \hline
\end{tabular}
\par\end{centering}
\caption{\label{tab:fit_result} The best-fit values with $1\sigma$
and $2\sigma$ errors for $\Omega_{m0},\, h$ and $\Omega_{b0}$ in the
ghost dark energy model.}

\end{table}

\begin{table}
\begin{centering}
\begin{tabular}{|c|c|c|c|}
\hline
parameter  & $\Omega_{m0}$  & $h$  & $\Omega_{b0}$\tabularnewline
\hline
\hline
best-fit  & $0.273$  & $0.703$  & $0.045$\tabularnewline
\hline
\end{tabular}
\par\end{centering}

\caption{\label{tab:lcdm}The best-fit values for the $\Lambda CDM$
model, using the same data sets.}

\end{table}

With the best-fit value of $\Omega_{m0}=0.257$, the transition
between the dust phase and the de Sitter phase occurs at
$a_{*}=\left(4\frac{\Omega_{m0}}{\Omega_{DE}^{2}}\right)^{\frac{1}{3}}=1.23$.
The universe begins to accelerate at
$a_{acc}=\frac{a_{*}}{2}=0.615$, or in terms of redshift,
$z_{acc}=0.625$. And the present EoS of DE
$\omega_{0}=-\frac{1}{\Omega_{m0}+1}=-0.796$.

$\chi^{2}$ of best-fit value of this model is
$\chi_{min}^{2}=607.192$ for $dof=575$. The reduced $\chi^{2}$
equals to $1.056$ which is acceptable. But $\chi_{min}^{2}$ is
larger than the one for the $\Lambda CDM$ model, $\chi_{\Lambda
CDM}^{2}=554.264$. A similar conclusion is also reached by other
authors using different data set~\cite{Basilakos:2009wi}.

It is not hard to understand why in this model, $\chi_{min}^{2}$ is
large, compared to the $\Lambda CDM$ model. Recently many model
independent studies on dynamics of DE show that current
observational data favor $\omega\sim-1$, at least at low redshift
\cite{Chevallier:2000qy}. But from (\ref{eq:w_Omegam}) we can see
that $\omega_{0}\sim-1$ requires a small $\Omega_{m0}$
($\Omega_{m0}\sim0$), which goes against CMB and BAO observation. As
a result, when we combine these different observational data sets to
do joint likelihood analysis, the final $\chi^{2}$ becomes large.

%%%%%%%%%%%%%%%%%%%%%%%%%%%%%%%%%%%%%%%%%%%%%%%%%%%%%%%%%%%%%%%%%%%%%%%%%%%%

\section{Adiabatic Sound Speed and Interaction}

In this section, we will return to the simplified model introduced
in section 2, and further study this model, where there are only two
components, DE and CDM in a flat universe.

\begin{figure}
\begin{centering}
\includegraphics{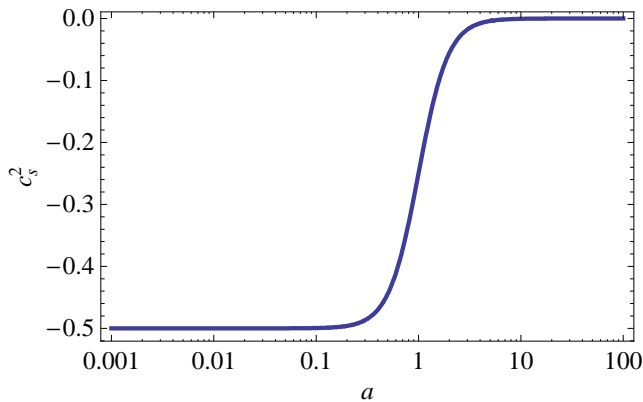}
\par\end{centering}

\caption{$c_{s}^{2}\sim a$ where $a_{*}=1$\label{fig:sound}}

\end{figure}

The squared adiabatic sound speed of the ghost DE model is found to
be
\begin{eqnarray*} c_{s}^{2} & = &
\frac{\dot{p}}{\dot{\rho}_{DE}}=-\frac{1}{2}\frac{1}{\left(\frac{a}{a_{*}}\right)^{3}+1}<0.\end{eqnarray*}
The evolution behavior of the  squared adiabatic sound speed is
shown in Figure \ref{fig:sound}. One can see from the figure that
$c_{s}^{2}$ leaps from $-\frac{1}{2}$ to $0$ at the time $a_{*}$.
Before $a_{*}$, $c_{s}^{2}$ is less than zero, but after
$a_{*}\sim1$, $c_{s}^{2}$ is approximate to zero. However,  it is
always negative!

Negative squared adiabatic sound speed may cause some problems. For
example, if a fluid evolves adiabatically and has no interaction
with gravity or other fluids, negative squared sound speed implies
an instability under perturbations. To see this clearly, let us
consider a Newtonian argument. Making use of Euler equation,
continuity equation of fluid and Poisson equation in Newtonian
gravity, we have \cite{book}
\[\ddot{\delta}+2H\dot{\delta}+c_{s}^{2}k^{2}\delta-4\pi G\bar{\rho}\delta=S_Q+S_{in},\]
where $\delta$ is energy density perturbation of the fluid. The
terms on the r.h.s are source terms. The first term $S_Q$ represents
interaction with other fluid, while the last term $S_{in}$ stands
for intrinsic entropy perturbation in the fluid itself. The $4\pi G$
term on the l.h.s represents the effect of gravity. If there are no
source terms, the negative $c_{s}^{2}$ term leads to an instability.
But the presence of source terms will change $\delta$'s behavior and
makes the thing complicated.

In the ghost DE model, DE interacts with gravity only and there is
no $S_Q$ or $S_{in}$ term. Thus the squared sound speed criterion
mentioned above shows DE will be unstable under perturbation in
Newtonian limit, where we have neglected relativistic effects.
However in a full relativistic treatment to discuss the stability of
this  DE model under linear perturbation, we need to define gauge
invariant variables and solve perturbed Einstein's equation and
conservation equations. However, it is beyond the scope of this
note.

From the above discussion, we see that negative squared adiabatic
sound speed may cause a potential instability. But it can be
improved if the source terms are present~\cite{Zimdahl:2010ae}.
Therefore in the rest of this section, we introduce the direct
interaction between DE and CDM and study the evolution dynamics of
the model.

In this case, Friedman equation still reads\begin{eqnarray*} H^{2} &
= & \frac{8\pi G}{3}\left(\alpha H+\rho_{m}\right)\end{eqnarray*}
 and conservation equations are modified to be \begin{eqnarray*}
\dot{\rho}_{m}+3H\rho_{m} & = & Q,\\
\dot{\rho}_{DE}+3(1+\omega)H\rho_{DE} & = & -Q,
\end{eqnarray*}
 where $Q$ denotes the interaction between DE and CDM. Since a
 generic form of $Q$ is not available, we consider three forms which
 are often discussed in the literature:
  $Q=3\bar{\alpha}H\rho_{DE},3\bar{\beta}H\rho_{m}$ and $3\bar{\gamma}H\rho_{tot}$,
where $\bar \alpha, \ \bar\beta$, and $\bar\gamma$ are three
constants, (see e.g. \cite{Wei:2007ut} for more references).  These
forms imply that energy transfers in a Hubble time is proportional
to energy density of DE, CDM and DE+CDM, respectively, and that
energy  transfers from DE to CDM if
$\bar{\alpha},\bar{\beta},\bar{\gamma}>0$, and vice versa.

In terms of dimensionless quantities, we have~\begin{eqnarray}
1 & = & \Omega_{DE}+\Omega_{m},\label{eq:1}\\
\frac{8\pi G}{3H^{2}}Q & = & \dot{\Omega}_{m}+2\frac{\dot{H}}{H}\Omega_{m}+3H\Omega_{m},\label{eq:2}\\
-\frac{8\pi G}{3H^{2}}Q & = &
\dot{\Omega}_{DE}+2\frac{\dot{H}}{H}\Omega_{DE}+3(1+\omega)H\Omega_{DE},
\label{eq:3}\end{eqnarray}
 where $\Omega_{DE}=\frac{8\pi G}{3H^{2}}\rho_{DE},\,\Omega_{m}=\frac{8\pi G}{3H^{2}}\rho_{m}$
is dimensionless energy density of DE and CDM, respectively. In
addition, by use of the linear relation between $\rho_{DE}$ and $H$,
we have
\begin{eqnarray} H\Omega_{DE} & = &
H_{0}\Omega_{DE0}=\text{const}.\label{eq:H_OmegaDE}\end{eqnarray}
Combining (\ref{eq:1}) and (\ref{eq:2}), one can get\begin{equation}
-\dot{\Omega}_{DE}+2\frac{\dot{H}}{H}\left(1-\Omega_{DE}\right)+3H\left(1-\Omega_{DE}\right)=\frac{8\pi
G}{3H^{2}}Q,\label{eq:2'}\end{equation} Putting (\ref{eq:3}) to
eliminate $Q$, we arrive at \begin{equation} 2\dot{H}+3\omega
H^{2}\Omega_{DE}+3H^{2}=0. \label{eq:4}\end{equation} Then from
(\ref{eq:2'}) and (\ref{eq:H_OmegaDE}), one can have the equation of
motion of $\Omega_{DE}$ as \begin{equation}
-\dot{\Omega}_{DE}\frac{2-\Omega_{DE}}{\Omega_{DE}}+3H_{0}\Omega_{DE0}\frac{1-\Omega_{DE}}{\Omega_{DE}}=\frac{8\pi
G}{3H^{2}}Q\equiv
H_{0}\Omega_{DE0}\Omega_{Q},\label{eq:EOM}\end{equation}
 where \[
\Omega_{Q}=\left\{ \begin{array}{ll}
3\bar{\alpha}, & {\rm when}\ Q=3\bar{\alpha}H\rho_{DE}\\
3\bar{\beta}\frac{1-\Omega_{DE}}{\Omega_{DE}}, &{\rm when}\ Q=3\bar{\beta}H\rho_{m}\\
3\bar{\gamma}\frac{1}{\Omega_{DE}}, & {\rm when}\
Q=3\bar{\gamma}H\rho_{tot}.\end{array}\right.\]
 Expressing this equation in terms of efolding-number $N\equiv\ln a$, and making use of
$\frac{d\Omega_{DE}}{dt}=\frac{d\Omega_{DE}}{dN}H_{0}\Omega_{DE0}/\Omega_{DE}$,
we obtain~\begin{equation}
-\Omega'_{DE}\frac{2-\Omega_{DE}}{\Omega_{DE}^{2}}+3\frac{1-\Omega_{DE}}{\Omega_{DE}}
=\Omega_{Q}.\label{eq:EOM_efold}\end{equation} Using
(\ref{eq:H_OmegaDE}),(\ref{eq:4}) and (\ref{eq:EOM}), we can get the
EoS of the DE
\begin{equation}
\omega=-\frac{1}{2-\Omega_{DE}}-\frac{2}{3}\frac{\Omega_{Q}}{2-\Omega_{DE}},
\label{eq:w_inter}\end{equation} while the EoS of the total fluid is
\begin{equation}
\omega_{tot}=-1-\frac{2}{3}\frac{\dot{H}}{H^{2}}=\omega\Omega_{DE},
\label{eq:wtot_inter}\end{equation} in the second equality, we have
used (\ref{eq:4}).  In addition, the deceleration parameter is given
by\begin{eqnarray} q & \equiv &
-\frac{\ddot{a}a}{\dot{a}^{2}}=\frac{1-2\Omega_{DE}}{2-\Omega_{DE}}-\frac{\Omega_{Q}\Omega_{DE}}
{2-\Omega_{DE}},
\label{eq:q_inter}
\end{eqnarray}
and the squared speed of sound reads
\begin{eqnarray} c_{s}^{2} & =
& \frac{\dot{p}}{\dot{\rho}_{DE}} =
\left(\Omega_{DE}\frac{d}{d\Omega_{DE}}-1\right)\frac{1+\frac{2}{3}\Omega_{Q}}
{2-\Omega_{DE}}.\label{eq:cs_inter}\end{eqnarray} We will solve
$\Omega_{DE},\,\omega,\,\omega_{tot}$ and $c_{s}^{2}$ analytically
for each $Q$ in the following. However, for the sake of briefness,
we will  discuss the case of $Q=3\bar{\alpha}H\rho_{DE}$ only in
some detail.

\subsection{$Q=3\bar{\alpha}H\rho_{DE}$}

In this case, (\ref{eq:EOM_efold}) becomes\[
-\Omega'_{DE}\left(2-\Omega_{DE}\right)+3\Omega_{DE}-3\left(\bar{\alpha}+1\right)\Omega_{DE}^{2}=0.\]
 Its solution is\begin{eqnarray}
3N+C & = &
2\ln\Omega_{DE}-\frac{1+2\bar{\alpha}}{1+\bar{\alpha}}\ln\left|1-
\left(\bar{\alpha}+1\right)\Omega_{DE}\right|,\label{eq:rhoDE_alpha}\end{eqnarray}
 where the integration constant $C=2\ln\Omega_{DE0}-\frac{1+2\bar{\alpha}}{1+\bar{\alpha}}
 \ln\left|1-\left(\bar{\alpha}+1\right)\Omega_{DE0}\right|$,
$\Omega_{DE0}$ is the dimensionless energy density of DE at present.

If $\bar{\alpha}<0$, from the solution (\ref{eq:rhoDE_alpha}), we
can see that $\Omega_{DE}$ can be larger than $1$ at late time which
is unphysical. This unphysical result comes from the fact that the
assumption on energy transfer rate $Q\propto\rho_{DE}$ is
oversimplified. For $\bar{\alpha}<0$, DE will gain energy from CDM
and energy density of CDM will become less and less. At some time,
$\rho_{m}$ becomes zero, and it is impossible to keep on
transferring energy to DE. But due to the oversimplified assumption
on $Q$, CDM will continue to lose energy which is incorrect
physically. Therefore, we will presume $\alpha>0$ in this
subsection.

\begin{figure}
\begin{centering}
\includegraphics{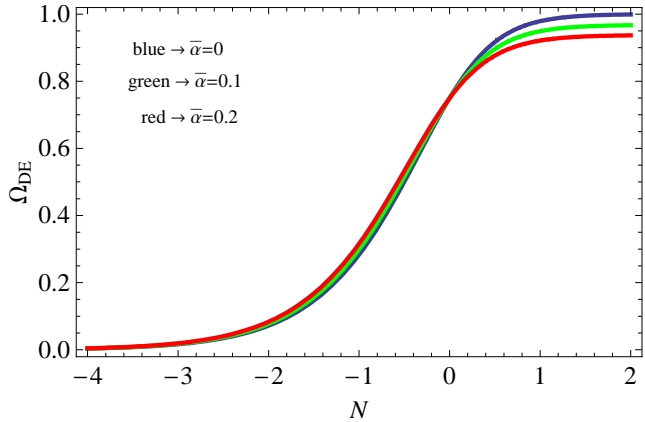}
\par\end{centering}

\caption{$\Omega_{DE}\sim N$ where $N$ is efolding-number. Here
$\Omega_{DE0}=0.75$. The blue, green and red curves correspond to
$\bar{\alpha}=0,\,0.1$ and $0.2$,
respectively.\label{fig:omega_N_alpha}}

\end{figure}

From (\ref{eq:rhoDE_alpha}) one has $(\bar{\alpha}+1)\Omega_{DE}<1$.
It means that for $\bar{\alpha}>0$, $\Omega_{DE}$ will tend to a
constant $1/(1+\bar{\alpha})$, rather than 1. The relation of
$\Omega_{DE}\sim N$ is shown in Figure \ref{fig:omega_N_alpha}, it
also indicates that when $\bar{\alpha}$ is larger, the evolution of
$\Omega_{DE}$ will be flatter since more energy are injected into
CDM. Of course there is a upper limit for $\bar{\alpha}$, as
$\Omega_{DE}$ must be able to reach its present value
$\Omega_{DE0}\sim0.75$. For $\bar{\alpha}>0$, the coincidence
problem can be alleviated excellently, and if $\bar{\alpha}\sim
-1+1/\Omega_{DE0}$, this problem is completely solved. The similar
situation occurs as $Q\sim\rho_{tot}$ (see section 4.3).

\begin{figure}
\begin{centering}
\includegraphics{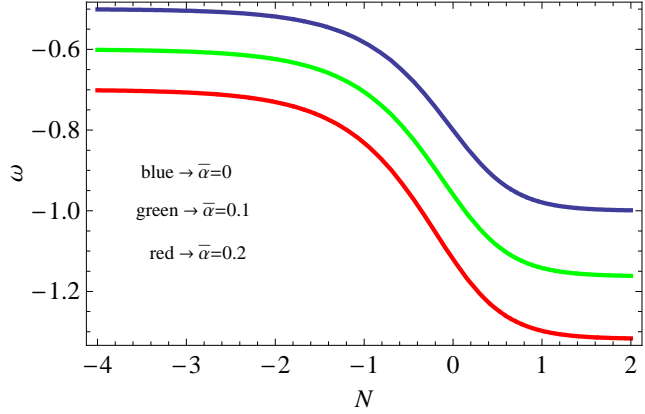}
\par\end{centering}

\caption{$\omega\sim N$ where $N$ is efolding-number. Here
$\Omega_{DE0}=0.75$. The blue, green and red curves correspond to
$\bar{\alpha}=0,\,0.1$ and $0.2$,
respectively.\label{fig:w_N_alpha}}

\end{figure}

In this case, the equation of state of DE
 is\[
\omega=-\frac{1+2\bar{\alpha}}{2-\Omega_{DE}}.\]
 We plot the relation $\omega\sim N$ in Figure \ref{fig:w_N_alpha}.
We have shown in section 2, it is impossible for $\omega$ to cross
phantom divide without interaction. Nevertheless from this figure we
can see that the situation is changed with the help of interaction
term. $\omega$ will cross $-1$ from the quintessence regime to
phantom regime and approach to $-1-2\bar{\alpha}$ at late time when
$\bar{\alpha}\ne0$.

\begin{figure}
\begin{centering}
\includegraphics{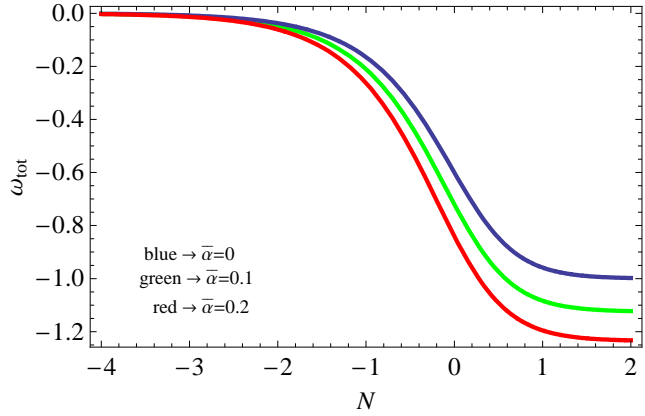}
\par\end{centering}

\caption{$\omega_{tot}\sim N$ where $N$ is efolding-number. Here
$\Omega_{DE0}=0.75$. The blue, green and red curves correspond to
$\bar{\alpha}=0,\,0.1$ and $0.2$, respectively.
\label{fig:wtot_N_alpha}}

\end{figure}

The equation of state  of the total fluid is\[
\omega_{tot}=-\left(1+2\bar{\alpha}\right)\frac{\Omega_{DE}}{2-\Omega_{DE}}\]
 and it is plotted in Figure \ref{fig:wtot_N_alpha}.
As we expect again, $\omega_{tot}$ can be smaller than $-1$ in the
presence of interaction term and reaches its asymptotic value
$-1-2\bar{\alpha}<-1$ at late time. Therefore, in this  model the
universe will end with the big rip singularity in the
future~\cite{r24,r25}. And from this figure, we can see that the
universe begins to accelerate earlier with larger $\bar{\alpha}$.
Note that according to the calculations in section 2 and section 3,
the universe begins to accelerate at $z_{acc,\bar{\alpha}=0}=0.6$ in
the case without interaction. Thus the acceleration of the universe
occurs at $z_{acc,\bar{\alpha}}>0.6$ when the interaction is present
if one keeps $\Omega_{m0}=0.257$.

Finally, the squared speed of sound reads\begin{eqnarray} c_{s}^{2}
& = &
\frac{\dot{p}}{\dot{\rho}_{DE}}=-2\left(1+2\bar{\alpha}\right)\frac{1-\Omega_{DE}}
{\left(2-\Omega_{DE}\right)^{2}},\label{eq:cs_alpha}\end{eqnarray}
 the squared sound speed is always smaller than $0$ because
we assumed $\bar{\alpha}>0$. This means that this form of
interaction cannot make $c_s^2$ positive.

\subsection{$Q=3\bar{\beta}H\rho_{m}$}

\begin{figure}
\begin{centering}
\includegraphics{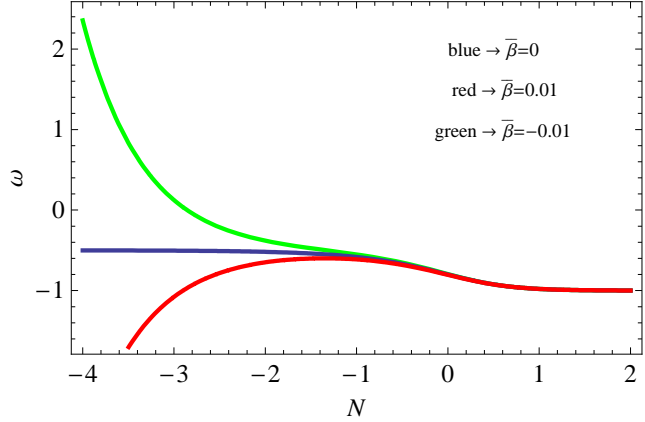}
\par\end{centering}

\caption{$\omega\sim N$ where $N$ is efolding-number. Here
$\Omega_{DE0}=0.75$. Blue, green and red curves correspond to
$\bar{\beta}=0,\,-0.01$ and $0.01$,
respectively.\label{fig:w_N_beta}}

\end{figure}

In this case, (\ref{eq:EOM_efold}) becomes

\[
-(2-\Omega_{DE})\Omega'_{DE}+3(1-\bar{\beta})\Omega_{DE}(1-\Omega_{DE})=0.\]
 its analytical solution reads\[
3\left(1-\beta\right)N+C=2\ln\Omega_{DE}-\ln\left(1-\Omega_{DE}\right),\]
 where the integration constant
 $C=2\ln\Omega_{DE0}-\ln\left(1-\Omega_{DE0}\right)$.
 We can see from the analytical solution that $\Omega_{DE}$ varies
from $0$ at early time to $1$ at late time. The EoS  of DE is
\[
\omega=-\frac{1}{2-\Omega_{DE}}-2\bar{\beta}\frac{1}{2-\Omega_{DE}}\frac{1-\Omega_{DE}}{\Omega_{DE}}.\]
 One can see  in this case that at early time $\omega\rightarrow-\bar{\beta}/\Omega_{DE}$
as $\Omega_{DE}\rightarrow0$. The plot is shown in Figure
\ref{fig:w_N_beta}. If $\bar{\beta}>0$, $\omega$ will increase from
$-\infty$ to some local maximum at some point and then decrease to
its asymptotic value $-1$ at late time. Unlike the situation we have
discussed in subsection 4.1, $\omega$ will cross $-1$ from phantom
regime to quintessence regime in this case. If $\bar{\beta}<0$,
$\omega$ will decay monotonically from $+\infty$ to $-1$ and never
cross $-1$. However, no matter what the value of $\bar{\beta}$ is,
the late-time asymptotic behaviors of $\omega$ are all the same.

The EoS of the total fluid is \begin{eqnarray*} \omega_{tot} & = &
-\frac{2\bar{\beta}+\left(1-2\bar{\beta}\right)\Omega_{DE}}{2-\Omega_{DE}}=\left\{
\begin{array}{ll}
-\bar{\beta} & \textrm{  at early time}\\
-1 & \textrm{  at late time.}\end{array}\right.\end{eqnarray*}
 The reasonable value of $\bar{\beta}$ should be $\left|\bar{\beta}\right|<1$.
Thus the big rip singularity will be avoided in this case.

Finally, we give the squared speed of sound

\begin{eqnarray*}
c_{s}^{2} & = &
-2\frac{1-\Omega_{DE}}{\left(2-\Omega_{DE}\right)^{2}}-2\bar{\beta}
\left[\frac{4-5\Omega_{DE}+2\Omega_{DE}^{2}}{\left(2-\Omega_{DE}\right)^{2}\Omega_{DE}}\right],
\end{eqnarray*}
 from which we can find that $c_{s}^{2}\rightarrow-2\bar{\beta}/\Omega_{DE}$
at early time. Therefore, this interaction term will not make
$c_{s}^{2}$ positive as well if $\bar \beta >0$.  Furthermore, at
early time the speed of sound will be larger than speed of light if
$\bar{\beta}<0$.

\subsection{$Q=3\bar{\gamma}H\rho_{tot}$}

In this case (\ref{eq:EOM_efold}) becomes

\[
-\Omega'_{DE}\frac{2-\Omega_{DE}}{\Omega_{DE}^{2}}+3\frac{1-\Omega_{DE}}{\Omega_{DE}}
=3\bar{\gamma}\frac{1}{\Omega_{DE}}.\]
 Its analytical solution reads \[
3N+C=\frac{2}{1-\bar{\gamma}}\ln\Omega_{DE}-\left(\frac{1+\bar{\gamma}}{1-\bar{\gamma}}\right)
\ln\left|1-\bar{\gamma}-\Omega_{DE}\right|,\]
 where $C=\frac{2}{1-\bar{\gamma}}\ln\Omega_{DE0}-\left(\frac{1
 +\bar{\gamma}}{1-\bar{\gamma}}\right)\ln\left|1-\bar{\gamma}-\Omega_{DE0}\right|.$
The EoS of DE\[
\omega=-\frac{1}{2-\Omega_{DE}}-2\frac{\bar{\gamma}}{2-\Omega_{DE}}\frac{1}{\Omega_{DE}}.\]
Once again $\omega$ will diverge at early time when the interaction
is present.  The EoS of the total fluid is
\[
\omega_{tot}=-\frac{2\bar{\gamma}}{2-\Omega_{DE}}-\frac{\Omega_{DE}}{2-\Omega_{DE}},\]
and the squared speed of sound reads
\begin{eqnarray*}
c_{s}^{2} & = &
-2\frac{1-\Omega_{DE}}{\left(2-\Omega_{DE}\right)^{2}}
+2\bar{\gamma}\frac{3\Omega_{DE}-4}{\Omega_{DE}\left(2-\Omega_{DE}\right)^{2}}\end{eqnarray*}
which is also divergent at early time $\Omega_{DE}\to 0$.

%%%%%%%%%%%%%%%%%%%%%%%%%%%%%%%%%%%%%%%%%%%%%%%%%%%%%%%%%%%%%%%%%%%%%%%%%%%%5

\section{Conclusion and Discussion}

In this note we investigated a DE model whose energy density is
proportional to Hubble parameter with a coefficient which is roughly
order of $\Lambda_{QCD}^{3}$. It  gives the right order of magnitude
of observed energy density of DE. We studied its cosmological
evolution. In this DE model, the universe has a de Sitter phase at
late time and begins to accelerate at redshift around
$z_{acc}\sim0.6$.

We also fitted this model with observational data including SnIa,
BAO, CMB, BBN and Hubble parameter data. The best-fit values of
parameters of the model are $\Omega_{m0}=0.257,\,
h=0.662,\,\Omega_{b0}=0.054$. However, the minimal $\chi^{2}$ gives
$\chi_{min}^{2}=607.192$, while in the $\Lambda CDM$ model,
$\chi_{\Lambda CDM}^{2}=554.264$ for the same data sets. Namely the
simple $\chi^2$ analysis seemingly implies that current data do not
favor the ghost DE model, compared to the $\Lambda CDM$ model.
Clearly this result is not conclusive, further study is needed.

We also found that the squared sound speed of the DE is negative,
which may give rise to a potential instability of the model under
perturbation. We further studied the cosmological dynamics of the
model by considering there exists some interaction between DE and
CDM. Three kinds of interaction forms are discussed. In all cases,
the negative squared sound speed is still there with a proper
coefficient for the interaction terms. Clearly the potential
instability should be studied seriously by investigating linearized
Einstein equations, not just calculating the squared sound speed of
the DE, which is currently under investigation. If the instability
indeed exists, then the ghost DE model has to be further modified or
to be abandoned.

%%%%%%%%%%%%%%%%%%%%%%%%%%%%%%%%%%%%%%%%%%%%%%%%%%%%%%%%%%%%%%%%%%%%%%%

\begin{acknowledgments}
RGC thanks N. Ohta for various valuable discussions on the QCD
ghost. HBZ thanks Bin Hu for helpful discussions. This work was
supported in part by the National Natural Science Foundation of
China (No. 10821504, No. 10975168 and No.11035008), and  by the
Ministry of Science and Technology of China under Grant No.
2010CB833004, and also by a grant from the Chinese Academy of
Sciences.

\end{acknowledgments}

%%%%%%%%%%%%%%%%%%%%%%%%%%%%%%%%%%%%%%%%%%%%%%%%%%%%%%%%%%%%%%%%%%%%%%
%%%%%%%%%%%%%%%%%% Refer %%%%%%%%%%%%%%%%%%%%%%%%%%%%%%%%%%%%%%%%%%%%%
%%%%%%%%%%%%%%%%%%%%%%%%%%%%%%%%%%%%%%%%%%%%%%%%%%%%%%%%%%%%%%%%%%%%%%

\vspace*{0.2cm}

%\cite{Riess:1998cb}

\end{document}